\documentstyle [12pt,epsf] {article}

\newcommand{\req}[1]{(\ref{#1})}

\newcommand{\nwc}{\newcommand}
\nwc{\btu}{\bigtriangleup}
\nwc{\cd}{\cdot}
\nwc{\zd}{{\bf Z}$_3$\ }
\nwc{\hyp} {\hyphenation}
\hyp{orbi-fold}
\hyp{theo-ries}
\hyp{theo-ry}
\hyp{regu-lari-zation}
\newcommand{\Z}{\ZZ}
\def\bfone{\relax{\rm 1\kern-.35em 1}}
\def\inbar{\vrule height1.5ex width.4pt depth0pt}
\def\IC{\relax\,\hbox{$\inbar\kern-.3em{\mss C}$}}
\def\ID{\relax{\rm I\kern-.18em D}}
\def\IF{\relax{\rm I\kern-.18em F}}
\def\IH{\relax{\rm I\kern-.18em H}}
\def\II{\relax{\rm I\kern-.17em I}}
\def\IN{\relax{\rm I\kern-.18em N}}
\def\IP{\relax{\rm I\kern-.18em P}}
\def\IQ{\relax\,\hbox{$\inbar\kern-.3em{\rm Q}$}}
\def\IR{\relax{\rm I\kern-.18em R}}
\def\ZZ{\relax{\hbox{\mss Z\kern-.42em Z}}}
\font\cmss=cmss10 \font\cmsss=cmss10 at 7pt
\def\ZZ{\relax\ifmmode\mathchoice
{\hbox{\cmss Z\kern-.4em Z}}{\hbox{\cmss Z\kern-.4em Z}}
{\lower.9pt\hbox{\cmsss Z\kern-.4em Z}}
{\lower1.2pt\hbox{\cmsss Z\kern-.4em Z}}\else{\cmss Z\kern-.4em Z}\fi}
\nwc{\ten}{ten--dimensional}
\nwc{\four}{four--dimensional}
\nwc{\gev} {{\rm GeV}}
\nwc{\tev} {{\rm TeV}}
\nwc{\mP} {$M_{\rm Planck}$}
\nwc{\mx} {$M_{\rm X}$}
\nwc{\ms} {$M_{\rm string}$}
\nwc{\sieb}{\mbox{\boldmath $\ov{27}$}}
\nwc{\sie}{\mbox{\boldmath ${27}$}}
\nwc{\lag}{Lagrangian}

\newcommand{\tcgc}{threshold corrections to the gauge couplings}
\newcommand{\tc}{threshold corrections}

\nwc{\be}  {\begin{equation}}
\nwc{\ee}  {\end{equation}}
\nwc{\ba}  {\begin{array}}
\nwc{\ea}  {\end{array}}
\nwc{\bdm} {\begin{displaymath}}
\nwc{\edm} {\end{displaymath}}
\nwc{\bea} {\be\ba{lcl}}
\nwc{\eea} {\ea\ee}
\nwc{\bda} {\bdm\ba{lcl}}
\nwc{\eda} {\ea\edm}
\nwc{\bc}  {\begin{center}}
\nwc{\ec}  {\end{center}}
\nwc{\ds}  {\displaystyle}
\nwc{\bmat}{\left(\ba}
\nwc{\emat}{\ea\right)}
\nwc{\nn} {\nonumber}
\nwc{\nnn} {\nonumber \vspace{.2cm} \\ }
\nwc{\ra}{\rightarrow}
\nwc{\lra}{\longrightarrow}
\nwc{\p} {\partial}

\nwc{\scr}  {\scriptstyle}
\nwc{\tx}  {\textstyle}
\nwc{\scs} {\scriptscriptstyle}
\nwc{\ov}  {\overline}
\nwc{\hb}  {\bar h}
\nwc{\xb}  {\bar x}
\nwc{\yb}  {\bar y}
\nwc{\zb}  {\bar z}
\nwc{\wb}  {\bar w}
\nwc{\Ob}  {\bar O}
\nwc{\Yb}  {\bar Y}
\nwc{\ep} {\epsilon}
\nwc{\de} {\delta}
\nwc{\Th} {\Theta}
\nwc{\th} {\theta}
\nwc{\al} {\alpha}
\nwc{\si} {\sigma}
\nwc{\Si} {\Sigma}
\nwc{\om} {\omega}
\nwc{\Om} {\Omega}
\nwc{\Ga} {\Gamma}
\nwc{\ga} {\gamma}
\nwc{\bet} {\beta}
\nwc{\La} {\Lambda}
\nwc{\la} {\lambda}
\nwc{\Sc}  {{\cal S}}
\nwc{\Rc}  {{\cal R}}
\nwc{\Dc}  {{\cal D}}
\nwc{\Oc}  {{\cal O}}
\nwc{\Cc}  {{\cal C}}
\nwc{\gc}  {{\cal g}}
\nwc{\Pc}  {{\cal P}}
\nwc{\Mc}  {{\cal M}}
\nwc{\Ec}  {{\cal E}}
\nwc{\Fc}  {{\cal F}}
\nwc{\Hc}  {{\cal H}}
\nwc{\Kc}  {{\cal K}}
\nwc{\Wc}  {{\cal W}}

\nwc{\Fcp} {{\cal F}^\pr}
\nwc{\Hcp} {{\cal H}^\pr}

\nwc{\Xc}  {{\cal X}}
\nwc{\Gc}  {{\cal G}}
\nwc{\Zc}  {{\cal Z}}
\nwc{\Nc}  {{\cal N}}

\nwc{\xc}  {{\cal x}}
\nwc{\Ac}  {{\cal A}}
\nwc{\Bc}  {{\cal B}}
\nwc{\Uc} {{\cal U}}
\nwc{\Vc} {{\cal V}}
\nwc{\Lc} {{\cal L}}
\nwc{\Qc} {{\cal Q}}

\nwc{\lng} {\langle}
\nwc{\rng} {\rangle}

\nwc{\lf} {\left}
\nwc{\ri} {\right}

\nwc{\diag} {{\rm diag}}
\nwc{\inv}  {{\rm inv}}
\nwc{\mod}  {{\ \rm mod\ }}
\nwc{\dete}  {{\rm det}}
\nwc{\tr}  {{\rm tr}}
\nwc{\im}  {{\rm Im}}
\nwc{\re}  {{\rm Re}}

\nwc{\h} {\frac{1}{2}}
\nwc{\fc} {\frac}

\def\KK{\relax{\rm I\kern-.18em K}}
\def\RR{\relax{\rm I\kern-.18em R}}
\def\NN{\relax{\rm I\kern-.18em N}}
\def\PP{\relax{\rm I\kern-.18em P}}

\def\zz{\relax{\sf Z\kern-.3em Z}}
\def\ZZ{\relax{\sf Z\kern-.4em Z}}
\def\ZZZ{{\relax{\sf Z}\kern -.5em Z}}
\def\ZZZ{Z\kern -0.37em Z}
\def\QQ{{\rm \kern .25em
             \vrule height1.4ex depth-.12ex width.06em\kern-.31em Q}}
\def\CC{{\rm \kern .25em
             \vrule height1.4ex depth-.12ex width.06em\kern-.31em C}}
\if@twoside\oddsidemargin25mm
\else\oddsidemargin25mm\evensidemargin25mm\marginparwidth25mm\fi%
\begin{document}
\begin{titlepage}
{\sf
\begin{flushright}
{TUM--HEP--225/95}\\
{SFB--375/22}\\
{hep--th/9510009}\\
{October 1995}
\end{flushright}}
\vfill
\vspace{-1cm}
\begin{center}
{\large \bf How to Reach the Correct
$\mbox{\boldmath $sin$}^{\mbox{\boldmath $2$}}
\mbox{\boldmath $\th$}_{\mbox{\boldmath w}}$ and
$\mbox{\boldmath $\al$}_{\mbox{\boldmath s}}$
in String Theory$^{\mbox{\boldmath $\ast$}}$}

\vskip 1.2cm
{\sc H. P. Nilles$^{1,2}$} {\ \small \ and}
{\ \ \sc S. Stieberger$^{1,3}$}\\
\vskip 1.5cm
{\em $^1$Institut f\"{u}r Theoretische Physik} \\
{\em Physik Department} \\
{\em Technische Universit\"at M\"unchen} \\
{\em D--85747 Garching, FRG}
\vskip 1cm
{\em $^2$Max--Planck--Institut f\"ur Physik} \\
{\em ---Werner--Heisenberg--Institut---}\\
{\em D--80805 M\"{u}nchen, FRG}
\vskip 1cm
{and}\\
\vskip .6cm
{\em $^3$Institut de Physique Th\'eorique}\\
{\em Universit\'e de Neuch\^atel}\\
{\em CH--2000 Neuch\^atel, SWITZERLAND}
\end{center}
\vfill

\thispagestyle{empty}

\begin{abstract}
Effective theories with the matter content
of the minimal supersymmetric Standard Model below the
string scale $M_{\rm string}$ predict a wrong value
for the weak--mixing angle $\sin^2\th_{\rm W}$
and strong coupling constant $\al_{\rm S}$ at the scale $M_{\rm Z}$.
To resolve this problem one needs large
threshold corrections. At the same time
one would like to  avoid introducing new intermediate scales that are small
compared to $M_{\rm string}$.
Two requests which seem to be incompatible.
We show how both requirements can be satisfied
in a class of (0,2) heterotic superstring compactifications
with a natural choice of the vevs of the moduli fields entering the moduli
dependent string threshold corrections.
\end{abstract}

\vskip 5mm \vskip0.5cm
\hrule width 5.cm \vskip 1.mm
{\small\small  $^\ast$ Supported by the
 "Sonderforschungsbereich 375--95: Research in Particle--Astrophysics" of the
Deutsche Forschungsgemeinschaft and the EEC under contract no.
SC1--CT92--0789.}
\end{titlepage}

LEP and SLC high precision electroweak data predict
for the minimal supersymmetric Standard Model (MSSM)
with the lightest Higgs mass in the range $60 \gev < M_{\rm H}<150 \gev$

\be
\ba{rcl}
\sin^2 \hat \th_{\rm W}(M_{\rm Z})&=&0.2316\pm0.0003\\
\al_{em}(M_{\rm Z})^{-1}&=&127.9\pm0.1\\
\al_{\rm S}(M_{\rm Z})&=&0.12\pm0.01\\
m_t&=& 160^{+11+6}_{-12-5}\gev\ ,
\ea\label{exp}
\ee
for the central value $M_{\rm H}=M_{\rm Z}$ in the $\ov{ MS}$ scheme
\cite{langacker}.
This is in perfect agreement with the recent CDF/D0 measurements of $m_t$.
Taking the first three values as input parameters leads to gauge coupling
unification at $M_{\rm GUT}\sim 2\cdot 10^{16}\gev$ with $\al_{\rm GUT}\sim
\fc{1}{26}$ and
$M_{\rm SUSY}\sim 1 {\rm TeV}$ \cite{uni00,langacker}.
Slight modifications arise from light SUSY thresholds, i.e. the splitting
of the sparticle mass spectrum,
the variation of the mass of the second Higgs doublet
and two--loop effects. Whereas these
effects are rather mild, huge corrections may arise from heavy thresholds
due to mass splittings at the high scale $M_{heavy}\neq M_{\rm GUT}$
arising from the infinite many
massive string states \cite{lan93}.

In heterotic superstring theories all couplings are related to
the universal string coupling constant $g_{\rm string}$ at
the string scale $M_{\rm string}\sim \al'^{-1/2}$, with $\al'$ being the
inverse string tension. It is a free parameter which is fixed by the dilaton
vacuum expectation value $g_{\rm string}^{-2}=\fc{S+\ov S}{2}$.
In general this amounts to
string unification, i.e. at the string scale \ms\ all gauge
and Yukawa couplings are proportional
to the string coupling and are therefore related to each other.
For the gauge couplings (denoted by $g_a$) we have \cite{gin87}:

\be\label{hyper}
g^{2}_ak_a=g_{\rm string}^2=\fc{\kappa^2}{2\al'}\ .
\ee
Here, $k_a$ is
the Kac--Moody level of the group factor labeled by $a$.
The string coupling $g_{\rm string}$
is related to the gravitational coupling constant
$\kappa^{2}$. In particular this means that string
theory itself provides gauge coupling
and Yukawa coupling unification even in absence of
a grand unified gauge group.

To make contact with the observable world
one constructs the field--theoretical low--energy limit of a
string vacuum. This is achieved by integrating out
all the massive string modes corresponding to excited string states
as well as states with momentum
or winding quantum numbers in the internal dimensions.
The resulting theory then describes the
physics of the massless string excitations at low energies
$\mu \ll M_{\rm string}$ in field--theoretical
terms.
If one wants to state anything about higher energy scales one has to
take into account \tc\ $\triangle_a(M_{\rm string})$
to the bare couplings $g_a(M_{\rm string})$
due to the infinite tower of massive string modes. They change
the relations \req{hyper} to:

\be
g_a^{-2}=k_ag_{\rm string}^{-2}+\fc{1}{16\pi^2}\triangle_a\ ,
\label{triangle}
\ee
The corrections in \req{triangle}
may spoil the string tree--level result \req{hyper} and
split the one--loop gauge couplings at
$M_{\rm string}$.
This splitting could allow for an effective unification at a scale
$M_{\rm GUT} <M_{\rm string}$ or destroy the unification.

The general expression of $\triangle_a$ for heterotic tachyon--free
string vacua is given in \cite{vk}. Various contributions to $\triangle_a$
have been determined for several classes of models:
First in \cite{vk} for two $\Z_3$ orbifold models with a (2,2)
world--sheet supersymmetry \cite{dhvw}.
This has been extended to fermionic constructions
in \cite{fermionic}. Threshold corrections for (0,2) orbifold models
with quantized Wilson lines \cite{hpn1} have been
calculated in \cite{mns}.
Threshold corrections for the quintic threefold
and other Calabi--Yau manifolds \cite{chsw} with gauge group
$E_6\times E_8$ can be found in \cite{ber1,kl2}.
In toroidal orbifold compactifications ~\cite{dhvw}
moduli dependent threshold corrections
arise only from N=2 supersymmetric sectors. They have been
determined for some orbifold compactifications in
\cite{DKL2}--\cite{cfilq} and for
more general orbifolds in \cite{ms1}.
The full moduli dependence\footnote{A lowest expansion
result in the Wilson line modulus has been obtained in \cite{agnt2,clm2}.}
of threshold corrections
for (0,2) orbifold compactifications with continuous Wilson lines
has been first derived in \cite{ms4,ms5}.
These models contain continuous background gauge fields in addition
to the usual moduli fields \cite{hpn2}. In most of the cases
these models are (0,2) compactifications.
In all the above orbifold examples the threshold corrections $\triangle_a$
can be decomposed into three parts:

\be
\triangle_a=\tilde \triangle_a-b_a^{N=2}\triangle+k_a\ Y\ .
\label{form}
\ee
Here the gauge group dependent part is divided into two pieces:
The moduli independent part $\tilde \triangle_a$ containing
the contribution of the N=1 supersymmetric sectors as
well as scheme dependent parts which are proportional to $b_a$.
This prefactor $b_a$ is related to the one--loop
$\bet$--function: $\bet_a=b_ag_a^3/16\pi^2$.
Furthermore the moduli dependent part $b^{N=2}_a\triangle$ with
$b_a^{N=2}$ being related to the anomaly coefficient $b'_a$ by
$b_a^{N=2}=b_a'-k_a\de_{\rm GS}$.
The gauge group independent part $Y$ contains the gravitational
back--reaction to the background gauge fields as well as other universal parts
\cite{vk,dfkz,kl2,kk}. They are absorbed into the definition of
$g_{\rm string}$: $g_{\rm string}^{-2}=\fc{S+\ov S}{2}+\fc{1}{16\pi^2}Y$.
The scheme dependent parts are the
IR--regulators for both field-- and string theory as well as
the UV--regulator for field theory. The latter is put into the definition of
$M_{\rm string}$ in the $\ov{\rm DR}$ scheme \cite{vk}:

\be
M_{\rm string}=2\fc{e^{(1-\ga_{\rm E})/2} 3^{-3/4}}{\sqrt{2\pi \al'}}=
0.527\ g_{\rm string} \times 10^{18}\ {\rm GeV}\ .
\label{kaprel}
\ee
The constant of the string IR--regulator as well as
the universal part due to gravity were
recently determined in \cite{kk}.

The identities \req{triangle} are the key to extract any string--implication
for low--energy physics. They serve as boundary conditions for
our running field--theoretical couplings valid below \ms\ \cite{basicweinberg}.
Therefore they are the foundation of any discussion about both
low--energy predictions and gauge coupling unification.
The evolution equations\footnote{We neglect the N=1 part of
$\tilde \triangle_a$ which is small compared to
$b_a^{N=2}\triangle$ \cite{vk,fermionic,mns}.} valid below $M_{\rm string}$

\be
\fc{1}{g_a^2(\mu)}=\fc{k_a}{g_{\rm string}^2}+\fc{b_a}{16\pi^2}
\ln \fc{M_{\rm string}^2}{\mu^2}-\fc{1}{16\pi^2}b^{N=2}_a\triangle\ ,
\label{running}
\ee
allow us to determine $\sin^2 \th_{\rm W}$ and
$\al_{\rm S}$ at $M_Z$.
After eliminating $g_{\rm string}$ in the second and third equations
one obtains

\bea\label{mz}
\sin^2\th_{\rm
W}(M_Z)&=&\ds{\fc{k_2}{k_1+k_2}-\fc{k_1}{k_1+k_2}\fc{\al_{em}(M_Z)}{4\pi}
\lf[\Ac\ln\lf(\fc{M_{\rm string}^2}{M_Z^2}\ri)-\Ac'\ \triangle\ri]\ ,}\nnn
\al_{S}^{-1}(M_Z)&=&\ds{\fc{k_3}{k_1+k_2}\lf[\al_{em}^{-1}(M_Z)-\fc{1}{4\pi}\Bc
\ln\lf(\fc{M_{\rm string}^2}{M_Z^2}\ri)+\fc{1}{4\pi}\Bc'\ \triangle\ri]\ ,}
\eea
with  $\Ac=\fc{k_2}{k_1}b_1-b_2, \Bc=b_1+b_2-\fc{k_1+k_2}{k_3}b_3$ and
$\Ac',\Bc'$ are
obtained by exchanging $b_i\ra b_i'$. For the MSSM one has $\Ac=\fc{28}{5},
\Bc=20$.
However to arrive at the predictions of the MSSM \req{exp}
one needs huge string
threshold corrections $\triangle$ due to the large value of
\ms\ \ $(3/5k_1=k_2=k_3=1)$:

\be\label{tri}
\triangle=\fc{\Ac}{\Ac'}\lf[\ln\lf(\fc{M_{\rm string}^2}{M_{\rm
GUT}^2}\ri)+\fc{32\pi\de_{\sin^2\th_{\rm W}}}{5\Ac\al_{em}(M_Z)}\ri]\ .
\ee
At the same time, the N=2 spectrum of the underlying theory
encoded in $\Ac',\Bc'$ which enters the threshold corrections
has to fulfill the condition

\be\label{cond}
\fc{\Bc'}{\Ac'}=\fc{\Bc}{\Ac}\ \fc{\ln\lf(
\fc{M_{\rm string}^2}{M_{\rm GUT}^2}\ri)+\fc{32\pi}{3\Bc}
\de_{\al_{\rm S}^{-1}}}{\ln\lf(\fc{M_{\rm string}^2}{M_{\rm GUT}^2}\ri)+
\fc{32\pi}{5\Ac}\fc{\de_{\sin\th_{\rm W}^2}}{\al_{em}(M_Z)}}\ ,
\ee
with the uncertainties $\de$ appearing in \req{exp}. In addition
$\de$  may also contain SUSY thresholds.

For concreteness and as an illustration
let us take the $\Z_8$ orbifold example of \cite{IL} with
$\Ac'=-2,\Bc'=-6$ and $b_1'+b_2'=-10$.
It is one of the few orbifolds left over after
imposing the conditions on target--space duality anomaly cancellation
\cite{IL}.
To estimate the size of $\triangle$ one may take in eq. ~\req{kaprel}
$g_{\rm string} \sim 0.7$ corresponding to $\al_{\rm string}\sim\fc{1}{26}$,
i.e. $M_{\rm string}/M_{\rm GUT}\sim 20$.
Of course this is a rough estimate since
 $M_{\rm string}$ is determined by the first eq. of \req{running}
together with \req{kaprel}. Nevertheless, the qualitative picture does
not change.
Therefore to predict the correct low--energy parameter \req{mz}
eq. \req{tri} tells us that one needs threshold correction of considerable
size:

\be
-16.3\leq\triangle\leq-17.1\ .
\label{size}
\ee

The  construction of a realistic unified string model boils down to the
question of how to achieve thresholds of that size. To settle the question
we need explicit calculations within the given candidate string model.
There we can encounter various types of threshold effects. Some depend
continuously, others discretely on the values of the moduli fields.
For historic reasons we also have to distinguish between thresholds
that do or do not depend on Wilson lines.
The reason is the fact that the calculations
in the latter models are considerably simpler and for some  time were
the only available results. They were then used to estimate the thresholds
in models with gauge group $SU(3)\times SU(2)\times U(1)$ and three
families, although as a string model no such orbifold can be constructed
without Wilson lines.
Therefore, the really relevant thresholds are, of course, the ones found
in the (0,2) orbifold models with Wilson lines \cite{ms4}
which may both break the gauge group and reduce its rank.
We will discuss the various contributions within the
framework of our illustrative model.
However the discussion can easily applied for all other orbifolds.
The threshold corrections depend on the $T$ and $U$ modulus
describing the size and shape of the internal torus lattice. In addition
they may depend on non--trivial gauge background fields encoded
in the Wilson line modulus $B$.

Moduli dependent threshold corrections $\triangle$ can be
of significant size for an appropriate
choice  of the vevs of the background fields $T,U,B,\ldots$
which enter these functions. Of course in the decompactification
limit $T\ra i \infty$ these corrections become always arbitrarily huge.
This is in contrast to fermionic string compactifications
or N=1 sectors of heterotic superstring compactifications.
There one can argue that {\em moduli--independent}
threshold corrections cannot become huge at all \cite{df}. This
is in precise agreement with the results found earlier in
\cite{vk,fermionic}.
In field theory threshold corrections can be estimated with the
formula \cite{basicweinberg}

\be\label{field}
\triangle=\sum_{n,m,k}\ln\lf(\fc{M_{n,m,k}^2}{M_{\rm string}^2}\ri)\ ,
\ee
with $n,m$ being the winding and momentum, respectively and $k$
the gauge quantum number of all particles running in the loop.
The string mass in the $N=2$ sector of the $\Z_8$ model we consider
later with a non--trivial gauge background in the internal
directions is determined by \cite{ms5} :

\bea\label{mass}
\al'M_{n,m,k}^2&=&4|p_R|^2\nnn
p_R&=&\ds{\fc{1}{\sqrt{Y}}\lf[(\fc{T}{2\al'}U-B^2)n_2+\fc{T}{2\al'}n_1
-Um_1+m_2+Bk_2\ri]}\nnn
Y&=&\ds{-\fc{1}{2\al'}(T-\ov T)(U-\ov U)
+(B-\ov B)^2\ .}
\eea
In addition a physical state $|n,m,k,l\rng$ has to obey the modular
invariance condition
$m_1n_1+ m_2n_2+ k_1^2-k_1k_2+k_2^2-k_2k_3-k_2k_4+k_3^2+k_4^2
=1-N_L-\h l^2_{E_8'}$. Therefore the sum in \req{field}
should be restricted to these states. This also guarantees
its convergence after a proper regularization.
In \req{field} cancellations between the contributions of
various string states may arise. E.g. at the critical point $T=i=U$
where all masses appear in integers of \ms\ such cancellations occur.
They are the reason for the smallness of the corrections
calculated in \cite{vk,mns} and in all the fermionic models \cite{fermionic}.
Let us investigate this in more detail.
The simplest case ($B=0$) for moduli dependent \tcgc\  was derived
in \cite{DKL2} :

\be
\triangle(T,U)=\ln\lf[\fc{-iT+i\ov T}{2\al'}\lf|\eta\lf(\fc{T}{2\al'}
\ri)\ri|^4\ri]+\ln\lf[(-iU+i\ov U)\lf|\eta(U)\ri|^4\ri]\ .
\label{22th}
\ee
Formula \req{22th} can be used for any toroidal orbifold compactifications,
where the two--dimensional subplane of the internal lattice
which is responsible for the N=2 structure factorize from the
remaining part of the lattice. If the latter condition
does not hold, \req{22th} is generalized \cite{ms1}.

\bdm
\ba{|c|c|c|c|c|c|}\hline
&&&&&\\[-.25cm]
\ &T/2\al'&U&M^2\al' &ln(M^2\al')&\Delta^{II}\\[-.25cm]
&&&&&\\ \hline &&&&&\\[-.25cm]
Ia&i&i     & 1  &0&-0.72\\[-.25cm]
&&&&&\\ \hline &&&&&\\[-.25cm]
Ib&1.25i&i&\fc{4}{5} &-0.22  &-0.76\\[-.25cm]
&&&&&\\ \hline &&&&&\\[-.25cm]
Ic&4.5i&4.5i      &\fc{4}{81} &-3.01  &-5.03\\[-.25cm]
&&&&&\\ \hline &&&&&\\[-.25cm]
Id&18.7i&i&\fc{10}{187} &-2.93&-16.3\\[-.25cm]
&&&&&\\ \hline
\ea
\edm
\begin{center}
{\em Table 1: Lowest mass $M^2$ of particles charged\\ under $G_A$
and threshold corrections $\triangle(T,U)$.}
\end{center}

In Table 1 we determine the mass of the lowest massive string state being
charged under the considered unbroken gauge group $G_A$
and the threshold corrections $\triangle(T,U)$ for some values
of $T$ and $U$.

The influence of moduli dependent \tc\ to low--energy physics [entailed
in eqs. \req{mz}]
has  until now only been discussed  for orbifold compactifications
without Wilson lines by using \req{22th}.
In these cases the corrections only depend on the two moduli $T,U$.
However to obtain corrections of the size $\triangle\sim-16.3$ one would need
the vevs $\fc{T}{2\al'}=18.7, U=i$
which are unnaturally far away from the
self--dual points \cite{ILR,IL}. It remains an open question whether and
how such big
vevs of $T$ can be obtained in a natural way in string theory.

A generalization of eq. \req{22th} appears when turning on non--vanishing
gauge background fields $B\neq0$.
According to \req{mass} the mass of the heavy string states now becomes
$B$--dependent and therefore also the \tc\ change.
This kind of corrections were recently determined in \cite{ms4}.
The general expression there is

\be\label{c12}
\triangle^{II}(T,U,B)=\fc{1}{12}\ln\lf[\fc{Y^{12}}{1728^4}
\lf|\Cc_{12}(\Om)\ri|^2\ri]\ ,
\ee
where $B$ is the Wilson line modulus,
$\Om=\lf(\ba{cc} \fc{T}{2\al'}&B\\ -B&U\ea\ri)$ and $\Cc_{12}$ is a combination
of $g=2$ elliptic theta functions explained in detail in \cite{ms5}.
It applies to gauge groups $G_A$
which are not affected by the Wilson line mechanism. The case where the gauge
group is broken by the Wilson line will be discussed later (those threshold
corrections will be singular in the limit of vanishing $B$).
Whereas the effect of quantized Wilson lines $B$ on \tc\ has already
been discussed  in \cite{mns} the function
$\Delta^{II}(T,U,B)$ now allows us to study the effect of a continuous
variation
in $B$.

\vspace{1cm}
\epsfbox[-80 0 500 210]{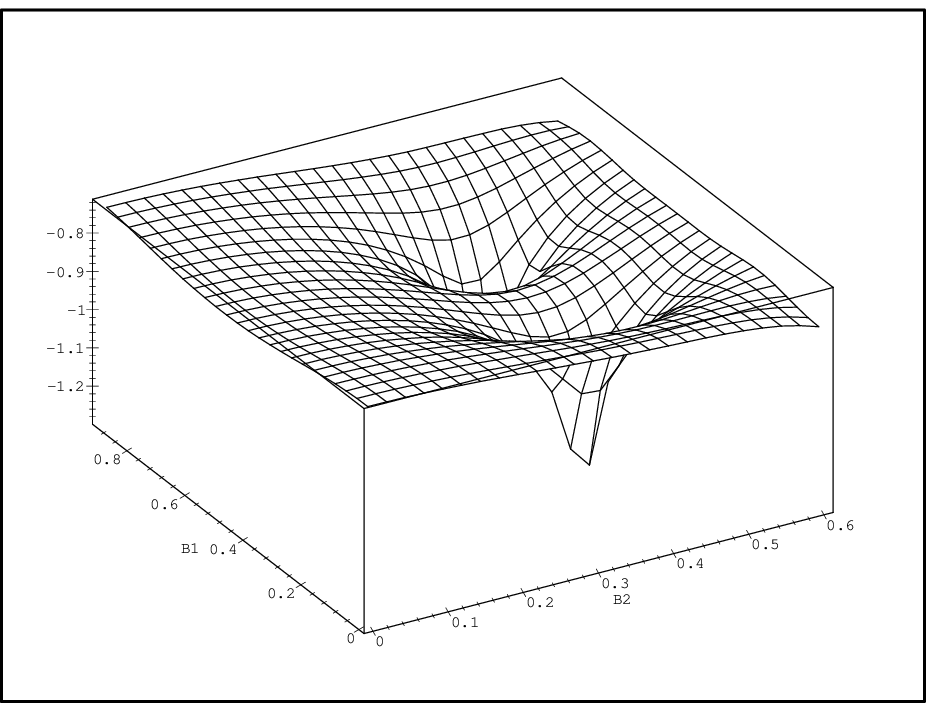}
\bc
{\small \em Fig.1 -- Dependence of the \tc\ $\triangle^{II}$\\
on the Wilson line modulus $B=B_1+iB_2$ for $\fc{T}{2\al'}=i=U$.}
\ec
We see in Fig.1  that the threshold corrections change very little
with the Wilson line modulus $B$. They are comparable
with $\triangle=-0.72$ corresponding to the case of $B=0$.
In this case eq. \req{c12} becomes eq. \req{22th}
for $\fc{T}{2\al'}=i=U$.

So far all these calculations have been done within models
where the considered gauge group $G_A$ is not broken by the Wilson line
and its matter representations are not projected out.
To arrive at SM like gauge groups with the matter content
of the MSSM one has to break the considered gauge group with
a Wilson line.

{}From the phenomenological point of
view \cite{wend}, the most promising
class of string vacua is provided by
(0,2) compactifications equipped with a non--trivial gauge background in the
internal space which breaks the $E_6$ gauge group
down to a SM--like gauge group \cite{w1,w2,hpn1,hpn2,stringgut}.
Since the internal space is not simply connected these gauge fields cannot be
gauged away and may break the gauge group.
Some of the problems  present in (2,2) compactifications
with $E_6$ as a  grand unified group like e.g.
the doublet--triplet splitting problem, the fine--tuning problem and
Yukawa coupling unification may be absent in (0,2) compactifications.
It is important that these properties can be studied in the full
string theory, not just in the field theoretic limit \cite{w1}.
The background gauge fields give rise to a new class of massless moduli
fields again denoted by $B$
which have quite different low--energy implications than the
usual moduli arising from the geometry of the internal manifold itself.
In this framework the question of string unification can now be discussed
for realistic string models.
The threshold corrections for our illustrative model take the form \cite{ms4}

\be\label{chi10}
\triangle^I(T,U,B)=\fc{1}{10}\ln \lf[Y^{10}\lf|\fc{1}{128}
\prod_{k=1}^{10}\vartheta_k(\Om)\ri|^4\ri]\ ,
\ee
where $\vartheta_k$ are the ten even
$g=2$ theta--functions \cite{ms5}. Equipped with this result we can now
investigate the influence of the B--modulus on the thresholds and see how the
conclusions of ref. \cite{ILR,IL} might be modified.
The results for a representative set of background vevs is displayed in
Fig.2.

\vspace{1cm}
\epsfbox[-80 0 500 210]{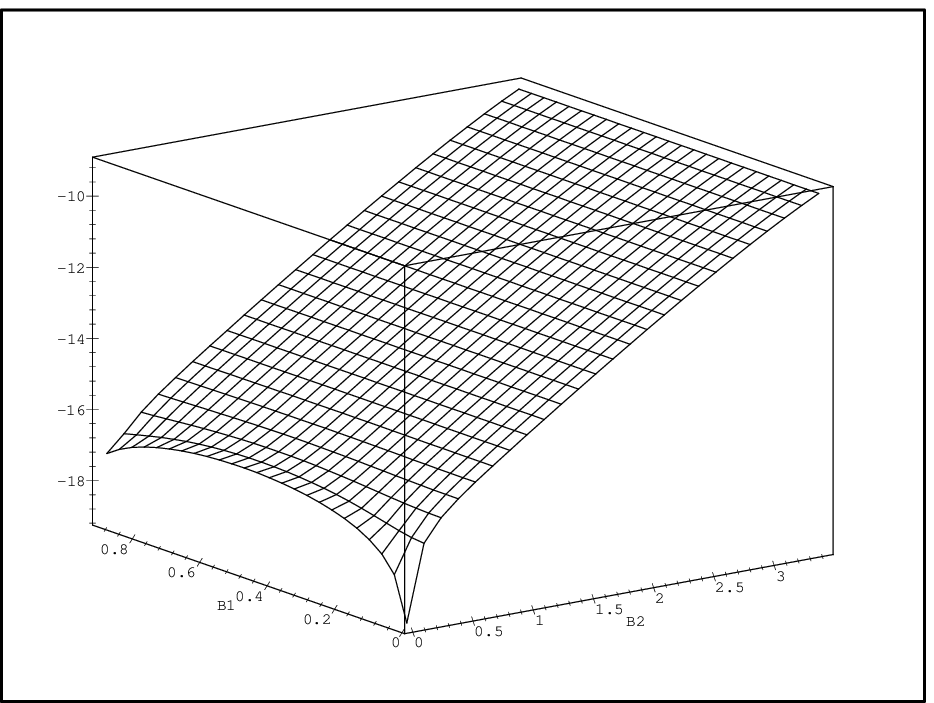}
\bc
{\small \em Fig.2 -- Dependence of the \tc\ $\triangle^I$ \\
on the Wilson line modulus $B=B_1+iB_2$ for $\fc{T}{2\al'}=4.5i=U$.}
\ec
{}From this picture we see that threshold corrections of
$\triangle\sim-16.3$ can be obtained for the choice of
$\fc{T}{2\al'}\sim 4.5i\sim U$ and $B=\h$. This has to be compared to the model
in ref. \cite{IL} where such a value was achieved with $T=18.7i$ and $B=0$.
This turns out to be a general property of the models under
consideration. With more moduli, sizeable threshold effects are achieved
even with moderate values of the vevs of the background fields.

The modulus plays the r\^ole of an adjoint Higgs field which breaks
e.g. the $G_A=E_6$ down to a SM like gauge group $G_a$.
According to eq. \req{mass} the vev
of this field gives some particles masses between zero and \ms.
This is known as the stringy Higgs effect.
Such additional intermediate fields may be very important to generate
high scale thresholds.
Sizeable \tc\ $\triangle$ can only appear if some particles have masses
 different from the string scale \ms\ and where not a cancellation
between different states as mentioned above takes place.
In particular some gauge bosons of $G_A$ become massive receiving the mass:

\be
\al'M_I^2=\fc{4}{Y}|B|^2\ .
\ee
As before let us investigate the masses of the lightest massive particles
charged under the gauge group $G_a$. For our concrete model we have
$M_{\rm string}=3.6\cdot 10^{17}\gev$.

\bdm
\ba{|c|c|c|c|c|c|c|}\hline
&&&&&&\\[-.25cm]
\ &T/2\al'&U&B&M_I\  [\gev] &ln(M_I^2\al') &\Delta^I\\[-.25cm]
&&&&&&\\ \hline &&&&&&\\[-.25cm]
IIa&i&i&\fc{1}{10^{5}}& 8.4\cdot 10^{12} &-23.0  &-10.03\\[-.25cm]
&&&&&&\\ \hline &&&&&&\\[-.25cm]
IIb&i&i&\h     &4.2\cdot 10^{17}     &-1.39           &-1.72\\[-.25cm]
&&&&&&\\ \hline &&&&&&\\[-.25cm]
IIc&1.25i&i&\h&3.7\cdot 10^{17}  &-1.61 &-2.12  \\[-.25cm]
&&&&&&\\ \hline &&&&&&\\[-.25cm]
IId&4i&i&\h     &2.1\cdot 10^{17} &-2.78    &-7.86\\[-.25cm]
&&&&&&\\ \hline &&&&&&\\[-.25cm]
IIe&4.5i&4.5i&\h     &9.3\cdot 10^{16}&-4.39   &     -16.3\\[-.25cm]
&&&&&&\\ \hline &&&&&&\\[-.25cm]
IIf&18.7i&i&\h&1.1\cdot 10^{16}&-4.31            &-43.3\\[-.25cm]
&&&&&&\\ \hline
\ea
\edm
\begin{center}
\vspace{.3cm}
{\em Table 2: Lowest mass $M_I$ of particles charged\\
under $G_a$ and threshold corrections $\triangle^I$ for $B\neq0$.}
\end{center}
\ \\
Whereas $\Delta^{II}$ describes \tc\ w.r.t. to a gauge group
which is not broken when turning on a vev of $B$,
now the gauge group is broken for $B\neq 0$ and in particular
this means that the threshold $\triangle^I$ shows a logarithmic singularity
for $B\ra 0$ when the full gauge symmetry is restored.
This behaviour is known from field theory and the effect from
the heavy string states
can be decoupled from the former:
Then the part of $\triangle_a^I$ in \req{triangle} which is only due to
the massive particles becomes \cite{ms4,clm2}

\be
\fc{b_A-b_a}{16\pi^2}\ln\fc{M_{\rm string}^2}{|B|^2}
-\fc{b_A'}{16\pi^2}\ln\lf|\eta\lf(\fc{T}{2\al'}\ri)\eta(U)\ri|^4\ ,
\ee
where the first part accounts for the new particles appearing
at the intermediate scale of $M_I$
and the other part takes into account the contributions
of the heavy string states.
One of the questions of string unification concerns the size of this
intermediate scale $M_I$.
In a standard grand unified model one would be tempted to identify
$M_I$ with $M_{\rm GUT}$. While this would also be a possibility for
string unification, we have in string theory in addition the possibility
to consider $M_I>M_{\rm GUT}$. The question remains whether the
thresholds in that case can be big enough, as we shall discuss in a moment.
Let us first discuss the general consequences of our
results for the idea of string unification without a grand unified
gauge group.
Due to the specific form of the threshold corrections
in eq. \req{running} unification always takes place if the condition
$\Ac\Bc'=\Ac'\Bc$
is met within the errors arising from the uncertainties in \req{exp}.
It guarantees that all three gauge couplings meet at
a single point $M_{\rm X}$ \cite{IL}:

\be
M_{\rm X}=M_{\rm string}\ e^{\h\fc{\Ac'}{\Ac}\triangle}\ .
\ee
For our concrete model this leads to $M_{\rm X}\sim2\cdot 10^{16}\gev$.
Given these results we can now study the relation between $M_I$ and
$M_{\rm X}$, which plays the r\^ole of the GUT--scale in string unified models.
As a concrete example, consider the model $IIe$ in Table 2. It leads
to an intermediate scale $M_I$ which is a factor 3.9 smaller than the
string scale, thus $\sim 10^{17}\gev$, although the apparent unification
scale is as low as $2\times 10^{16}\gev$. We thus have an explicit example
of a string model where all the non--MSSM particles are above
$9.3\cdot10^{16}\gev$, but still a correct prediction of the low energy
parameters
emerges.
Thus string unification can be achieved without the introduction of
a small intermediate scale.

Of course, there are also other possibilities which lead to the
correct low--energy predictions.
Instead of large threshold corrections one could consider
a non--standard hypercharge
normalization, i.e. a $k_1\neq 5/3$ \cite{ib}.
This would maintain gauge coupling unification at the string scale
{\em with} the correct values of $\sin^2\th_{\rm W}(M_Z)$ and
$\al_{\rm S}(M_Z)$.
However, it is very hard to construct such models.
A further possibility  would be to give up the idea of
gauge coupling unification within the MSSM by introducing
extra massless particles such as $\bf{(3,2)}$ w.r.t.
$SU(3)\times SU(2)$ in addition to those of the SM \cite{inter,df}.
A careful choice of these matter fields may lead to sizable
additional intermediate threshold corrections in \req{mz} thus
allowing for the correct low--energy data \req{exp}.
Unfortunately the price for that is exactly an introduction of a new
intermediate scale of $M_I\sim 10^{12-14}\gev$. It seems to be hard to
explain such a small scale naturally in the framework of string theory.
In some sense such a model can be compared to the model $IIa$ in table 2.
Other possible corrections to \req{mz} may arise
from an extended gauge structure between \mx\ and \ms.
However this might even enhance the disagreement
with the experiment \cite{df}. Finally a modification to \req{mz}
appears from the scheme conversion from the string-- or SUSY--based
$\ov{DR}$ scheme
 to the $\ov{MS}$ scheme relevant for the low--energy physics data \req{exp}.
However these effects are shown to be small \cite{df}.

Therefore we conclude with stating again the
new result that
string unification is easily achieved with moduli
dependent threshold corrections within (0,2) superstring compactification.
The Wilson line dependence
of these functions is comparable to that on the $T$ and $U$ fields thus
offering the interesting possibility of large thresholds with
background configurations of moderate size. All non--MSSM like
 states can be
heavier than $1/4$ of the string scale, still leading to an apparent
unification scale of $M_{\rm X}=\fc{1}{20}M_{\rm string}$.
We do not need vevs of the moduli fields that are of the order 20 away
from the natural scale, neither do we need to introduce particles at
a new intermediate scale that is small compared to $M_{\rm string}$.
The situation could be even more improved with a higher number of
moduli fields entering the threshold corrections:
They may come from other orbifold planes giving rise to N=2 sectors or
from additional Wilson lines.
We think that
the actual moderate vevs of the underlying moduli fields can be fixed
by non--perturbative effects as e.g. gaugino condensation.
\ \\ \\
{\em {\mbox{\boldmath $Acknowledgement:\ $}}}
We would like to thank Peter Mayr for helpful discussions
and comments and Alex Niemeyer for discussions and providing
a part of the computer program.


\end{document}